\documentclass{PoS}

\title{Direct Dark Matter Search with the CRESST II Experiment}

\ShortTitle{Search for Dark Matter with CRESST II}

\author{
         \speaker{J~Schieck$^{1,2}$},
	G~Angloher$^3$,
	A~Bento$^4$,
	C~Bucci$^5$,
	L~Canonica$^5$,
	X~Defay$^6$,
	A~Erb$^{6,7}$,
	F~v~Feilitzsch$^{6}$,
	N~Ferreiro~Iachellini$^3$,
	P~Gorla$^5$,
        A~G\"utlein$^{1,2}$,
	D~Hauff$^3$,
	J~Jochum$^8$,
	M~Kiefer$^3$,
	H~Kluck$^{1,2}$,
	H~Kraus$^9$,
	J-C~Lanfranchi$^6$,
	J~Loebell$^8$,
	M~Mancuso$^3$,
	A~M\"unster$^6$,
	C~Pagliarone$^5$,
	F~Petricca$^3$,
	W~Potzel$^6$,
	F~Pr\"obst$^3$,
	R~Puig$^{1,2}$,
	F~Reindl$^3$,
	K~Sch\"affner$^5$,
	S~Sch\"onert$^6$,
	W~Seidel$^3$,
	M~Stahlberg$^{1,2}$,
	L~Stodolsky$^3$,
	C~Strandhagen$^8$,
	R~Strauss$^3$,
	A~Tanzke$^3$,
	H~H~Trinh~Thi$^6$,
	C~T\"urko\v{g}lu$^{1,2}$,
	M~Uffinger$^8$,
	A~Ulrich$^6$,
	I~Usherov$^8$,
	S~Wawoczny$^6$,
	M~Willers$^6$,
	M~W\"ustrich$^3$
	and A~Z\"oller$^6$ \\
	\llap{$^1$}Institute for High energy Physics, Austrian Academy of Sciences, 1050 Vienna, Austria \\
	\llap{$^2$}Atominstitut, Technische Universit\"at Wien, 1040 Vienna, Austria \\
	\llap{$^3$}Max-Planck-Institut f\"ur Physik, D-80805 M\"unchen, Germany \\
	\llap{$^4$}Departamento de Fisica, Universidade de Coimbra, P3004 516 Coimbra, Portugal \\
	\llap{$^5$}INFN, Laboratori Nazionali del Gran Sasso, I-67010 Assergi, Italy \\
	\llap{$^6$}Physik-Department, Technische Universit\"at M\"unchen, D-85748 Garching, Germany \\
	\llap{$^7$}Walther-Mei\ss ner-Institut f\"ur Tieftemperaturforschung, D-85748 Garching, Germany \\
	\llap{$^8$}Eberhard-Karls-Universit\"at T\"ubingen, D-72076 T\"ubingen, Germany \\
	\llap{$^9$}Department of Physics, University of Oxford, Oxford OX1 3RH, United Kingdom \\}

\abstract{The quest for the particle nature of dark matter is one of the big open questions of modern physics. A well motivated candidate for dark matter is the so-called WIMP - a weakly interacting massive particle. Recently several theoretically well-motivated models with dark matter candidates in a mass region below the  WIMP mass-scale gained also a lot of interest, theoretically and experimentally.  The CRESST II experiment located at the Gran Sasso laboratory in Italy is optimised for the detection of the elastic scattering of these low-mass dark matter particles with ordinary matter. 
We show the results obtained with an improved detector setup with increased radio purity and enhanced background rejection and the results obtained with a dedicated low-threshold analysis of a single conventional detector module.
The limit achieved is the most stringent limit achieved for direct dark matter experiments in the mass region below 1.8 GeV/$c^{2}$. We will discuss the expected performance for new small CRESST-type detectors to be used during the next data taking phase. We conclude with an outlook of the future potential for direct dark matter detection using further improved CRESST CaWO$_{4}$ cryogenic detectors.}

\FullConference{38th International Conference on High Energy Physics\\
		3-10 August 2016\\
		Chicago, USA}
		
\begin{document}

\section{Introduction}
Several observations on different scales clearly point to the existence of dark matter. Up to now dark matter has been observed only via gravitational pull. 
The existence of a new particle, not being observed by previous and current experiments, offers the best explanation of dark matter. However,
the properties of this new particle are yet unknown. Several well motivated theories predict dark matter particle candidates in a large mass range from $\mu$eV/$c^{2}$
up to $\mathrm{M_{GUT}}$. These candidates are expected to interact with ordinary matter with an interaction strength well above the gravitational interaction scale, 
leading to a possible detection of scattering processes between dark matter and ordinary matter. \par
Recently dark matter candidates in the mass region of a few GeV/$c^{2}$ down to the sub-GeV/$c^{2}$ region (low-mass dark matter)
gained a lot of theoretical interest. This mass region can 
be still accessed with direct detection experiments, however, the experimental challenges differ from the one experienced by direct detection experiments 
looking for weakly interacting dark matter (WIMP), located in the mass region of tenth of GeV/$c^{2}$ to several TeV/$c^{2}$.
The expected differential scattering rate with respect to the recoil-energy falls exponentially and much more steeply for low-mass dark matter particles compared 
to particles with larger mass. Direct dark matter experiments looking for WIMPs require a large exposure in order to be sensitive to possible scattering processes. 
For direct detection experiments, searching for dark matter particles of a few GeV/$c^{2}$ or below, the detection energy threshold is crucial, while the
total exposure is of lower importance. \par
The CRESST experiment is a dedicated dark matter direct detection experiment searching for low-mass dark matter particles. In the following we will 
summarise the latest results obtained by CRESST II and published in \cite{Angloher:2014myn,Angloher:2015ewa,Angloher:2016jsl,Angloher:2016ktr}.
\section{The CRESST II Experiment}
\begin{figure}
\begin{center}
\includegraphics[width=.6\textwidth]{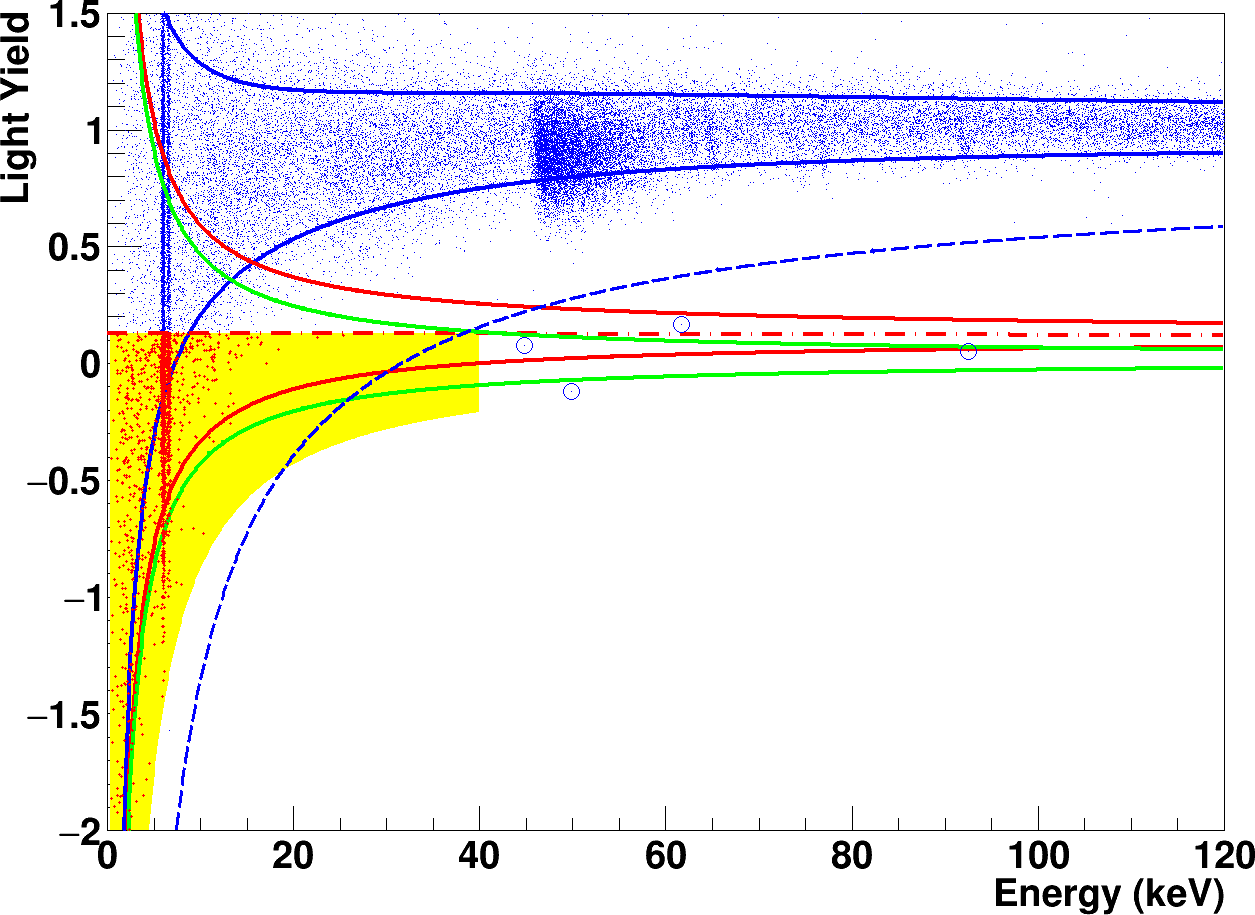}
\caption{Light yield as a function of the total recoil energy (phonon energy) for the Lise detector~\cite{Angloher:2015ewa}. The solid blue line 
indicates the upper and lower 90$\%$ boundaries for electron or photon scatters, the blue dashed line corresponds to the lower
 five $\sigma$ boundary.  The red (green)  lines correspond to the 90$\%$ boundaries for oxygen (tungsten) recoils. The yellow 
 box indicates the signal region. The four events indicated by a circle are statistically incompatible with leakage from the region of electron 
 or gamma scatters 
 and indicate the presence of another source of background events.}
\label{fig:LightYield}
\end{center}
\end{figure}

The CRESST II experiment is located at the underground laboratory {\it Laboratori Nazionali del Gran Sasso} in Italy. 
Dark matter particles are detected by looking for elastic scattering processes of dark matter with nuclei of CaWO$_{4}$ crystals. 
The experiment is operated at a temperature of about 15 mK. The total recoil energy deposited by the scattering process 
is measured by detecting the generated phonons, using a superconductor attached to the surface of the crystal and operated at its phase transition 
temperature. A second readout channel measures the scintillating light generated during the scattering process. Scattering of 
electrons or photons with the crystal produces significantly more light than dark matter scatters ({\it quenching}) and 
the light measurement allows the rejection of background events. 
For each event the light yield, defined as ratio of the energy measured in the light channel and the phonon 
channel, is determined. For electron or photon scatters the light yield is normalised to one, scatters of dark matter particles are
expected to have a lower light yield. For smaller recoil energies the uncertainty on the light yield increases and 
background events from electron or photon scattering events can potentially leak into the signal region.
The light yield versus phonon energy for data taken with a single detector is shown in Fig.~\ref{fig:LightYield}. The leakage 
of background type events into the signal region, indicated by the yellow box, is clearly visible. \par
The dominant background contribution originates from decays of crystal-intrinsic radioactivity. The overall sensitivity to
direct dark matter detection is limited by the number of unidentified background events in the region of interest. Recently
the CRESST collaboration started an in-house production of CaWO$_{4}$ crystals at the Technical University of Munich.
The radioactive contaminations are significantly reduced with a background rate as low as 3.5 counts /(kg\,keV\,d). 
\section{Results}
In this article we will mainly discuss results obtained with two different detector modules, called TUM40 and Lise. While the detector module 
TUM40~\cite{Angloher:2014myn} is built using a crystal from the in-house production with low intrinsic radioactive noise, the second detector 
is based on a conventional detection design and a commercial crystal with normal intrinsic background conditions~\cite{Angloher:2015ewa}.  
Data taken during a previous data taking period are used to test the Majorana character of neutrinos~\cite{Angloher:2016ktr}.
\subsection{Search for low-mass dark matter}
\label{StandAnalysis}
The key for detecting low-mass dark matter particles is the sensitivity for low energetic nuclear recoils. The expected differential  
dark matter-nucleus interaction rate increases exponentially towards low-mass dark matter particles. The analysis presented
here uses data from a single detector module. With the TUM40 detector module a detection threshold of about 600 eV for nuclear 
recoils is achieved, while for data taken with the Lise detector module the threshold is determined to be about 300 eV. The number
of selected events in the region of interest is used to set a limit on the dark matter-nucleon cross section. For the dark matter density 
and the dark matter velocity distribution standard astrophysical assumptions are used. A limit is set by assuming all events are originating
from dark matter scatters and the maximal possible cross-section is determined by using Yellin's optimal interval method, which is
considered to provide a conservative limit. \par
\begin{figure}
\begin{center}
\includegraphics[width=.8\textwidth]{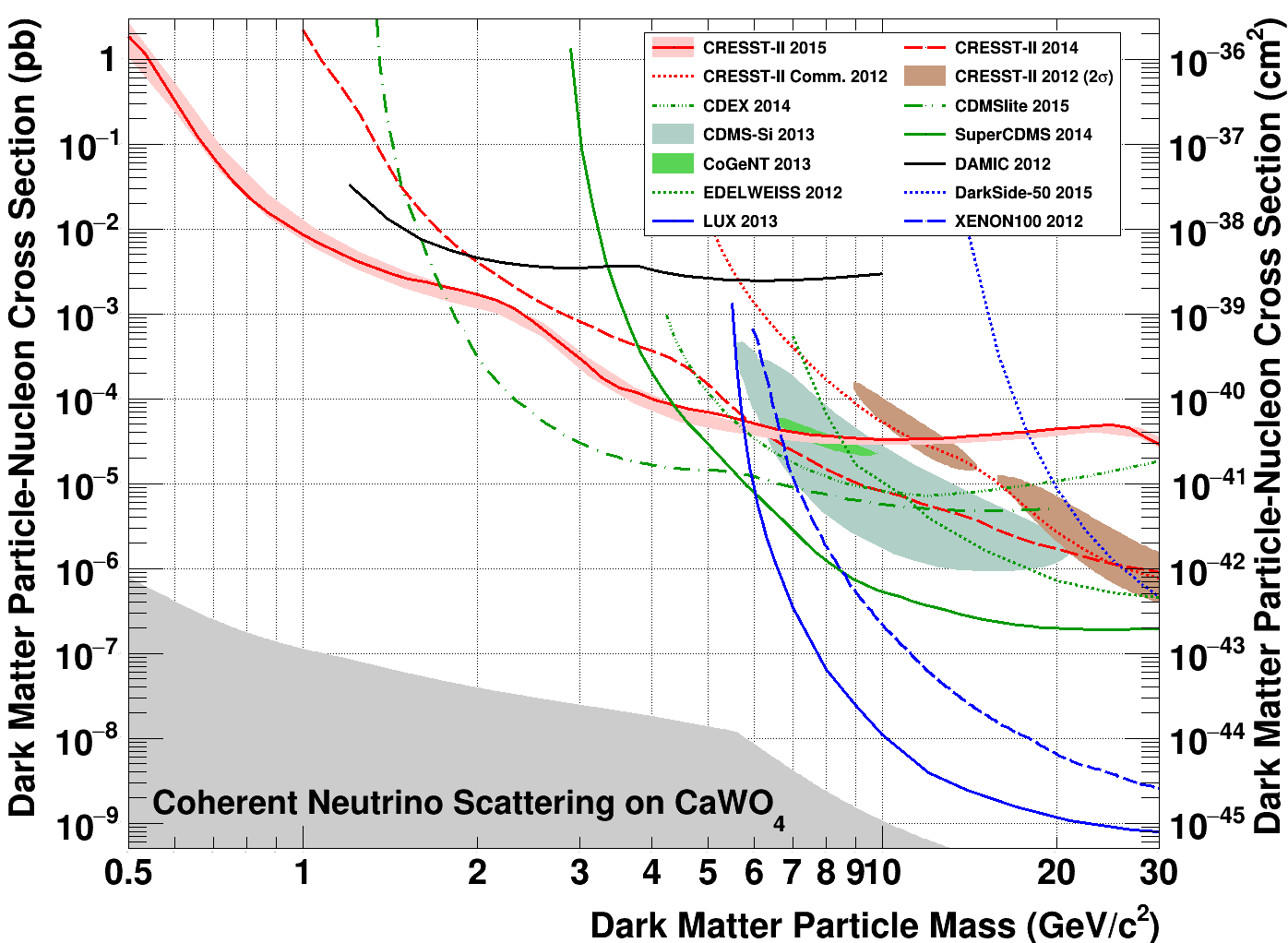}
\caption{Result of the dark matter searches using CRESST II detector modules~\cite{Angloher:2015ewa}. The red dashed line corresponds
to the limit obtained with the TUM40 detector module, the solid red line to the limit obtained with the Lise detector module. For further information
about the other results shown in the figure the reader is referred to the legend.}
\label{fig:ResultLowMassDM}
\end{center}
\end{figure}

The result is shown in Fig.~\ref{fig:ResultLowMassDM} and compared to other dark matter searches, using the same astrophysical input.
The Lise detector module with a lower energy threshold returns
the better limit  for lowest mass dark matter particles compared to the result obtained with the TUM40 detector module. 
For dark matter masses of several GeV/$c^{2}$ the limit
obtained by experiments based on liquid Xenon are better, since the overall exposure dominates the results. Due to 
higher background contamination the result of the Lise detector module is worse compared to the limit obtained with the TUM40 detector module.
Clearly for dark matter direct detection experiments CRESST II offers the best sensitivity for dark matter candidates below 1.8 GeV/$c^{2}$.
\subsection{Limit on the momentum dependent cross-section}
\label{MomAnalysis}
\begin{figure}
\begin{center}
\includegraphics[width=.6\textwidth]{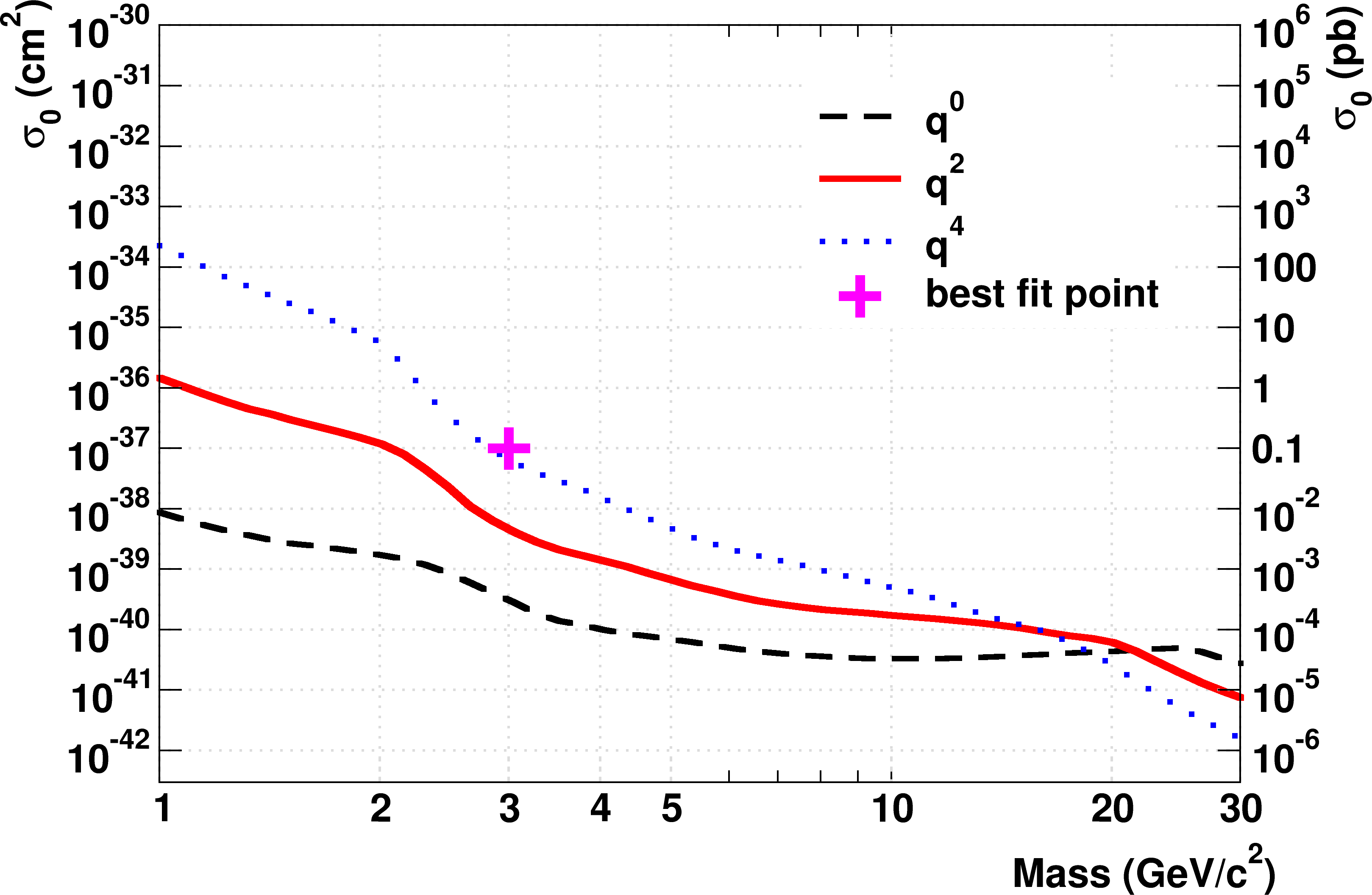}
\caption{The figure shows the limits for momentum dependent dark matter-nucleon cross sections using data
from the Lise detector module~\cite{Angloher:2015eza}. The exponent of the momentum dependent term is 
summarised in the legend. The best fit value, indicated by the cross, with an asymmetric dark matter model and a $q^{2}$ dependence, 
as obtained by~\cite{Vincent:2014jia}, is excluded.}
\label{fig:momentum}
\end{center}
\end{figure}

The results summarised in chapter~\ref{StandAnalysis} assume a momentum independent dark matter-nucleon cross section, leading to an
exponential increase of the differential event rate towards low recoil energies. Outstanding problems from comparing helioseismological predictions with solar models
can be solved by postulating asymmetric dark matter with a momentum dependent dark matter-nucleon cross section~\cite{Vincent:2014jia}. The standard 
cross-section is modified as 
\begin{math}
\sigma_{\mathrm{dark\,matter - nucleon}} = \sigma_{0} (q / q_{0})^{2},
\end{math}
with $q$ being the momentum transfer, $\sigma_{0}$ the standard momentum-independent cross-section and $q_{0}$ a reference value.
This cross-section behaviour leads to a decrease in the expected differential dark matter - nucleon cross section for lower recoil energies. 
For a momentum transfer of $q_{0}=40$ MeV, corresponding to a typical recoil
energy of 10 keV, the discrepancies between the helioseismological predictions
and the solar models can be resolved, assuming a dark matter particle with a mass of 3 GeV/$c^{2}$ and a cross section of $10^{-37}$ cm$^{2}$.
The data obtained with the Lise detector module can be reinterpreted assuming a momentum dependent cross-section~\cite{Angloher:2015ewa}. The limits corresponding to
different dependencies on the momentum transfer $q$ are summarised in Fig.~\ref{fig:momentum}. The best fit value discussed in~\cite{Vincent:2014jia} can be 
excluded with the recent CRESST II data set.
\subsection{Search for double electron capture processes}
In the Standard Model of particle physics neutrinos are the only possible particles which could act as Majorana-type particles. In this case the
particle would be it's own anti-particle. The observation of a simultaneous beta-decay of two nuclei, where the neutrino of one decay is 
absorbed by the other decay ($0\nu2\beta$), would be an unambiguous sign for the Majorana character of neutrinos. Besides this, 
the Majorana character can also be probed by a neutrino-less double electron capture process ($0\nu$EC), with
two electrons simultaneously captured and two protons transformed into neutrons, without emitting any neutrinos. 
The expected lifetime for such an process is much longer compared to $0\nu2\beta$ decays and any
possible signal is more difficult to observe. However, the current limits for the isotopes $^{40}$Ca and $^{180}$W are not very strong 
and can significantly be improved with the existing CRESST II data set. The double electron capture with two neutrinos 
released has also not been observed up to now and is part of this analysis as well. In this case the most probable decay
is based on the capture of two electrons from the K-shell ($2\nu2$K).  \par
Data from the CRESST II experiment taken between 2009 and 2011 are analysed to search for these processes~\cite{Angloher:2016ktr}. 
While for the results summarised in chapters ~\ref{StandAnalysis} and ~\ref{MomAnalysis} events with a light yield smaller than one
are used, here events consistent with an electron or gamma scatter and a light yield of around one are studied. For $0\nu$EC events the total energy is
absorbed in the detector, since no energy is carried away by any neutrino. A potential signal would show up as a peak 
for electron or gamma events with the expected energy release. No signal is observed and a limit is set for $0\nu$EC and $2\nu2$K decays 
for $^{40}$Ca and $^{180}$W isotopes. The limits determined for the different decays and isotopes are summarised in table~\ref{tab:BetaDecay}.
For the $2\nu2$K of $^{180}$W the limit improves by a factor of 30, compared to previous measurements, the other measurements 
improve the limit up to a factor of seven. 
\begin{table}
\begin{center}
\begin{tabular}{| c | c | c |}
\hline
Isotope & Process & T$_{1/2}$ [y] \\
\hline
$^{40}$Ca & $0\nu$EC & $> 1.40 \times 10^{22}$ \\ 
$^{40}$Ca & $2\nu2$K &  $> 9.92 \times 10^{21}$ \\ 
$^{180}$W & $0\nu$EC &  $> 9.39 \times 10^{18}$ \\ 
$^{180}$W & $2\nu2$K & $> 3.13 \times 10^{19}$ \\ 
\hline
\end{tabular}
\caption{The table summarises the limits for the lifetime T$_{1/2}$, obtained for the two isotopes $^{40}$Ca and $^{180}$W  and the two different decay 
scenarios $0\nu$EC and $2\nu2$K~\cite{Angloher:2016ktr}.}
\label{tab:BetaDecay}
\end{center}
\end{table}
\section{Outlook}
Recently a new data taking campaign with an improved CRESST II detector module design started.
The new design is based on smaller CaWO$_{4}$ crystals, leading to an even higher phonon density and
an increased sensitivity towards smaller nuclear recoil energies.  The design goal
is to reach an energy threshold of 100 eV. Modules are already installed in the cryostat and 
the detectors reached their operation temperature. \par
\begin{figure}
\begin{center}
\includegraphics[width=.8\textwidth]{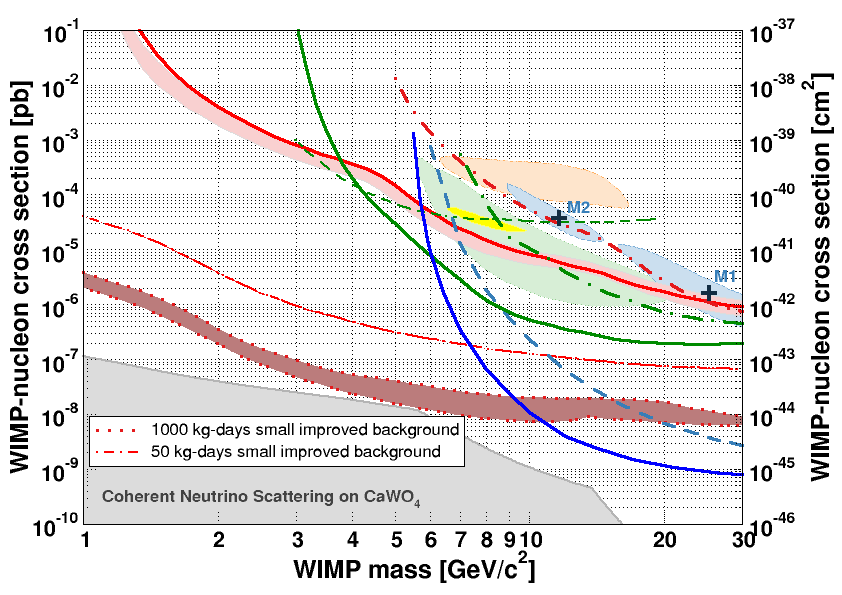}
\caption{Estimated sensitivity for dark matter detection using the CRESST experiment with an improved detector design with an expected
nuclear recoil energy threshold of 100 eV  and improved intrinsic radio purity by a factor of hundred ~\cite{Angloher:2015eza}. The dashed-dotted 
(dotted) line corresponds to an exposure of 50 kg-days (1000 kg-days). }
\label{fig:outlook}
\end{center}
\end{figure}

In a second operation phase the sensitivity will be further improved by lowering the intrinsic 
radioactive background of the crystals by a factor of 100. The expected sensitivity for
the the final operation phase is shown in Fig.~\ref{fig:outlook}. 
\section{Summary}
Several astrophysical measurements clearly indicate the existence of dark matter. The most likely scenario is a new type of particle.
Dedicated experiments search for 
elastic scattering of these dark matter particles with ordinary matter, CRESST II is one among them. The CRESST II experiment has the best 
sensitivity to detect low-mass dark matter particles in the masse region between 500 MeV/$c^{2}$ and 1.8 GeV/$c^{2}$. The CRESST collaboration
recently started an upgrade program to further improve the sensitivity in this mass region by several orders of magnitudes.

\end{document}